\documentclass[prl,aps,amssymb,twocolumn,showpacs,superscriptaddress]{revtex4}
\usepackage{amsfonts}
\usepackage{graphicx}

\begin{document}

\title{Doubled Full Shot Noise in Quantum Coherent Superconductor - Semiconductor Junctions}

\author{F. Lefloch, C. Hoffmann, M. Sanquer and D. Quirion}
 \altaffiliation[Present adress ]{ Max Planck Institut f\"ur Festk\"orper- \\ forschung, Heisenbergstr. 1, D-70569 Stuttgart, Germany.}
\affiliation{
D\'epartement de Recherche Fondamentale sur la Mati\`ere Condens\'ee/SPSMS, CEA Grenoble, 17 avenue des Martyrs, 38054 Grenoble Cedex 09, FRANCE.
}

\date{\today}

\begin{abstract}
We performed low temperature shot noise measurements in superconductor (TiN) - strongly disordered normal metal (heavily doped Si) weakly transparent junctions. We show that the conductance has a maximum due to coherent multiple Andreev reflections at low energy and that the shot noise is then {\it twice the Poisson noise ($S=4eI$)}. When the subgap conductance reaches its  minimum at finite voltage the shot noise changes to the normal value ($S=2eI$) due to a large quasiparticle contribution.
\end{abstract}

\maketitle

We know, from early measurements, that the full shot noise in an electronic device $S_{Poisson}=2qI$ (first measured by W. Schottky in a vacuum diode) is proportional to the mean value of the current $I$ and to the charge $q$ of the  carriers \cite{Schottky}. This result holds for N-I-N tunnel junctions where N is a normal metal and I an insulating barrier \cite{Blanter00} with $q=e$ the electronic charge. In S-I-N, due to electron pairing in the superconductor (S), the shot noise is expected to be twice the full shot noise : $S=4eI$. However, in such junctions the subgap current is very small and the shot noise is not measurable. The subgap current can be restored if the quasiparticles of the normal metal are coherently backscattered towards the interface. This reflectionless tunneling regime can be achieved by adding a second barrier in the normal part of the junction (S-I-N-I-N) or when the normal metal is disordered enough \cite{Quirion01,Kastalsky91,VanWees92,Quirion00,Beenakker97,Volkov94,Bakker,Magnee,Beltram}. The enhancement of the subgap current is only seen at low energy when the electron-hole coherence time in the normal metal is longer than the time it takes for a quasiparticles to return to the interface. Then, the coherent addition of two (or more) Andreev reflections, each of them with a very small probability $\Gamma ^{2}$ ($\Gamma$ is the transparency of the barrier), yields to an increase of the Andreev current through the interface. This effect can be large, leading to an Andreev current proportional to $\Gamma$ instead of $\Gamma ^{2}\ll \Gamma \ll 1$ and can be comparable to the normal current above the gap (also proportional to $\Gamma$). Another way to increase the subgap current is to use highly transparent S-N junctions. In this case doubled shot noise is predicted and has been observed experimentally at various temperatures \cite{Kozhevnikov00,Jehl00}. However, the noise level was always three times smaller than the doubled full shot noise because of the diffusive nature of the normal metal used in these experiments. Moreover, as shown experimentally and reproduced theoretically \cite{Nagaev01}, the doubling of the shot noise occurs at any energies below the superconducting gap and electron-hole coherence is not required.

\par In this letter, we report shot noise measurements in a junction where a superconductor (TiN) is in contact with heavily doped silicon.  We show that the shot noise is {\it twice the full shot noise at low energy} and equals the Poisson value at bias much smaller than the superconducting gap. This behavior evidences a crossover from a low bias Andreev dominated to a large bias quasiparticle dominated subgap conductance.
\par The sample is made of two 100/10 nm thick TiN/TiSi$_{2}$ contacts on top of a silicon substrate. The silicon is heavily doped ($n_{e}=2.10^{19}\, cm^{-3}$) over a thickness of $0.6\mu m$ \cite{Quirion00}. The two contacts are squares of 1 mm wide and the distance between them is $1\mu m$ or $2\mu m$. We present the results obtained with the $1\mu m$ sample and found the same results for the $2 \mu m$ sample. Because of the doping, the silicon is metallic ($k_{F}l_{e}\simeq 3$). Moreover, the superconducting TiN/TiSi$_{2}$ bilayer and the silicon are separated by a Schottky barrier which stays symmetric (non rectifying) in our voltage range. Therefore, the contact is described by a S-I-N junction where I stands for a tunnel barrier. The transport properties have been studied in details by Quirion at al. \cite{Quirion00}, and it was shown that the junction presents reflectionless tunneling behavior at low energy ($T \lesssim 250 mK$ and $V\lesssim 20\mu V$). 
Using reference \cite{Volkov94} for fitting the temperature dependence of the zero bias resistance, the following parameters have been obtained: the superconducting gap is $\Delta$ =0.22meV, the damping factor $\Gamma_{S}$ \cite{Dynes78} and the depairing rate $\gamma_{in}$ 
 are both relatively large: $\Gamma_{S}/ \Delta =0.15 \pm 0.01$ and $\gamma_{in}/ \Delta =0.27 \pm 0.05$. The typical Schottky barrier transparency is  $\Gamma =3.5 \ \ 10^{-2}$. The fitting of the subgap conductance implies many parameters and the quantitative agreement should be taken cautiously. As in previous reports of reflectionless tunneling \cite{Kastalsky91,Bakker,Magnee,Beltram} the large background subgap conductance cannot be attributed to Andreev reflection only, but to a large quasiparticule contribution due to a large $\Gamma_{S}$. Moreover most authors have concluded from conductance measurements to a non uniform barrier interface, that makes the theoretical comparison mostly qualitative. In this context noise measurements are necessary to discriminate between Andreev and quasiparticule contributions. 

We also know \cite{Quirion00} that coherent effects take place underneath the contacts and do not extend sidewise in the silicon between the contacts. The total sample can therefore be treated as two independent S-I-N junctions connected by a small piece of doped silicon of resistance $R_{Si}$. The sample resistance is :
\begin{equation}
R_{sample}=2R_{contact}+R_{Si}=2R_{contact}+N_{sq}R_{sq,Si}
\label{eq:resistance}
\end{equation}
where $R_{sq,Si}=24\Omega$ is the sheet resistance of the silicon and $N_{sq}=1/1000$ the number of squares between the two contacts. We also know that the current partially flows below the contacts before entering the superconductor. Therefore, $R_{contact}$ ($\simeq 0.4\Omega$ at low temperature) includes the resistance of the barrier and the resistance of the doped silicon underneath the superconductor which can be larger than in the native film. 
\par
\begin{figure}
\includegraphics[width=0.45\textwidth]{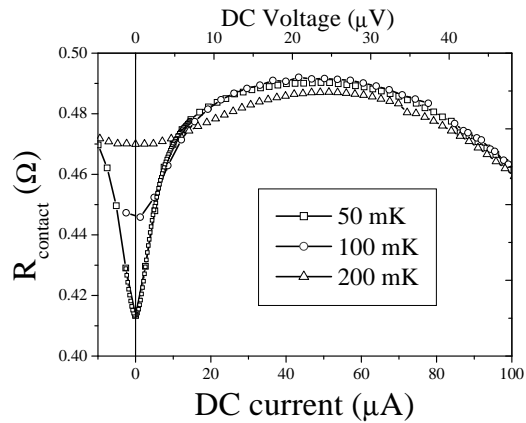}
\caption{Differential resistance at a contact for various temperatures. The upper scale is the voltage drop at the contact. The zero bias anomaly is characteristic for the coherent backscattering of quasiparticles in a S-I-N junction where N is a disordered metal (see text).}
\label{fig:noise1}
\end{figure}

 In figure \ref{fig:noise1}, we plotted the differential resistance of one contact as a function of the DC current and of the DC voltage drop at the contact, measured with the experimental set up used for the noise. We recover previous results which show that coherent effect appears at energies below $\simeq 20 \mu eV$, where the differential resistance shows a maximum \cite{Quirion00}. As long as the voltage stays above $20\mu V$, the differential resistance increases with decreasing energy because both the quasiparticles and the uncoherent Andreev contributions to the subgap current decrease \cite{BTK82}. For low DC voltage and low temperature, the coherence is established between successive reflections and the Andreev current increases.
From the temperature and voltage dependences of the  conductance it is possible to  evaluate the respective parts of the quasiparticle, uncoherent and coherent Andreev contributions. At voltage much larger than  $20\mu V$, both the Green's functions\cite{Volkov94} and the BTK\cite{BTK82} descriptions give a dominant quasiparticle contribution (the uncoherent Andreev contribution within the BTK model is less than 10 percents). This is  due to the large damping factor $\Gamma_{S}$ for the quasiparticles density of states and to an  elevated effective electron temperature $T_{eff}$ induced at finite voltage\cite{Quirion00}. Such phenomenological $T_{eff}$ is introduced to explain why  $eV_{max} \simeq k_{B}T_{max}$ where  $V_{max} \simeq 20\mu V$ is the voltage at which the resistance is maximum and  $T_{max} \simeq 250mK$ is the temperature at which the zero bias resistance is maximum. From the theory, such relation is not obeyed if one supposes that the  electron temperature is the base phonon temperature. At voltage smaller than  $20\mu V$, the Andreev contribution (treated within the Green's functions formalism \cite{Volkov94} which describes the coherent part) becomes larger than the quasiparticle part (below  $V=20\mu V$ the quasiparticle contribution is less than 10 percents). At  $V \simeq 40\mu V$, due to an  increase of $T_{eff}$,  both contributions are estimated  of the same order.

\par For shot noise measurements, we used a SQUID based experimental set up \cite{Jehl99} which measures the noise of a known macroscopic resistor ($R_{ref}=0.13\Omega$) in series with the sample. The total current noise power $S_{total}$ measured, is then given by :
\begin{equation}
S_{total}=2S_{contact}\frac{R^{2}_{contact}}{\left(\sum R\right)^{2}}+S_{Si}\frac{R^{2}_{Si}}{\left(\sum R\right)^{2}}+S_{ref}\frac{R^{2}_{ref}}{\left(\sum R\right)^{2}}
\label{eq:noise}
\end{equation}
Here, $S_{contact}$ is the noise we want to study, $S_{Si}$ the thermal noise generated by the silicon between the two contacts, $S_{ref}$ the thermal noise of the reference resistor and $\sum R$ is the sum of the resistances. We have chosen the sample to be such that $R_{contact} \gg R_{Si}, R_{ref}$, thus the total noise is principally the noise at the contact and the measurement is mainly insensitive to both the noise between the contacts and to the reference resistor noise. It is also insensitive to a change of the electronic temperature within these parts. At equilibrium, the noise is due to thermal fluctuations ($S=4k_{B}T/R$) and the total measured noise is $S=4k_{B}T/(\sum R)$. Far from equilibrium, equation (\ref{eq:noise}) applies if $R_{contact}$ is the differential resistance.

\par
\begin{figure}
\includegraphics[width=0.45\textwidth]{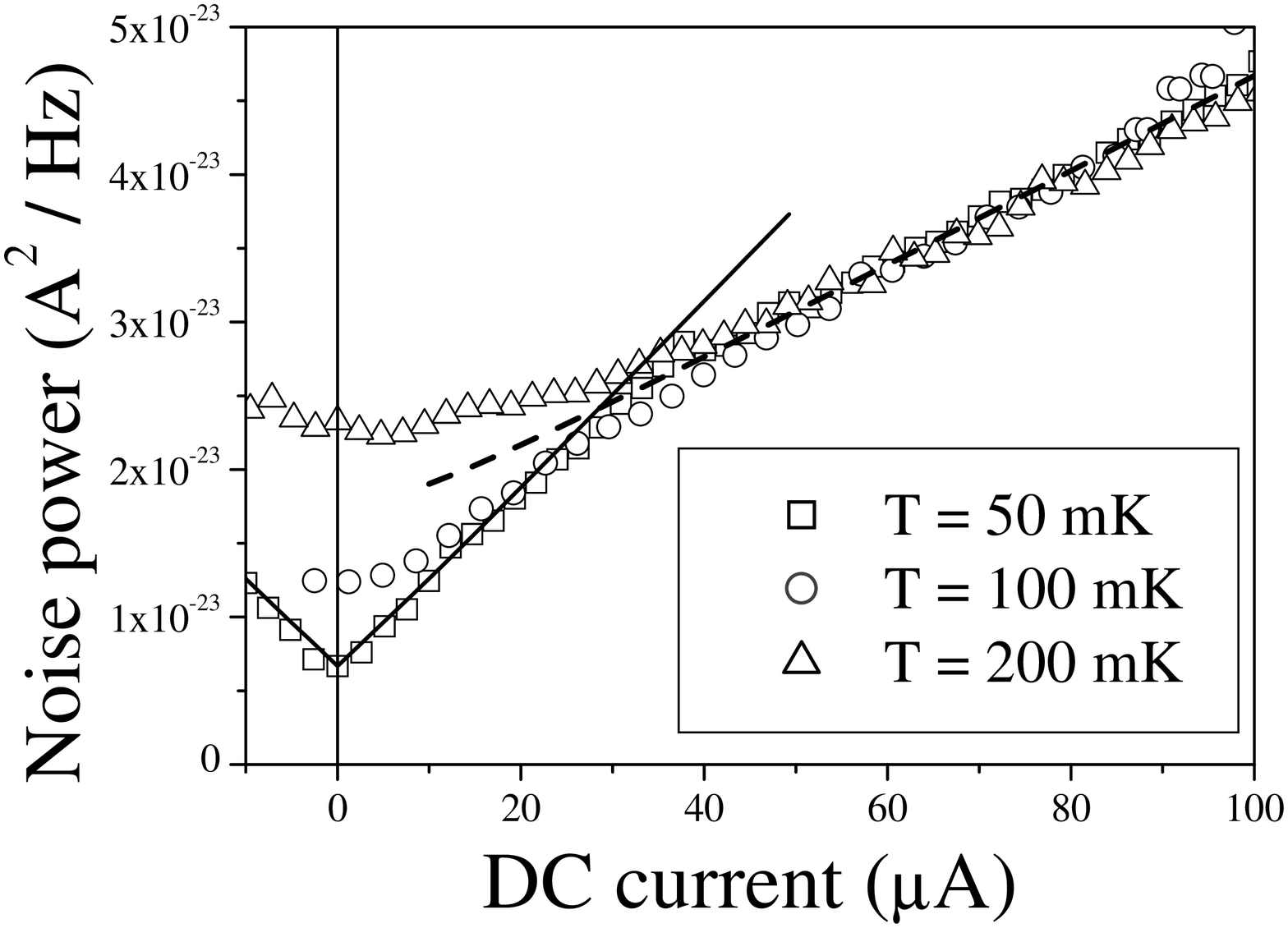}
\caption{Current noise power as a function of the average DC current. The solid and dashed lines correspond to $4e$ and $2e$ slopes respectively. The cross-over between the two behaviors occurs when the differential resistance shows a maximum (see figure \ref{fig:noise1}) indicating the change in the contributions to the current from  Andreev transport to  quasiparticles currents.}
\label{fig:noise2}
\end{figure}

 In figure \ref{fig:noise2}, we displayed the current noise power as a function of the DC current that passes through the contact. If we focus on the results obtained at $T = 50\,mK$, we clearly see that the shot noise exhibits a kink around $30 \mu A$ which corresponds to the value of current where the differential resistance shows a broad maximum (figure \ref{fig:noise1}).  In the low bias regime ($I<30\mu A$), the shot noise is proportional to the current with a slope $4e$ corresponding exactly to twice the full shot noise: this is our main result. Note that we do not see the thermal cross-over at $eV\simeq k_{B}T\simeq 5 \mu eV$ and that the solid line is $S=4eI+4k_{B}T/R_{contact}(V=0)$. At higher currents ($I>30\mu A$), the shot noise has a slope $2e$ as depicted by the dashed line in figure \ref{fig:noise2}. More precisely, the dashed line is : $S=2eI+1.5\,10^{-23}A^{2}/Hz=\frac{2}{R}(eV+E_{c})$ with $E_{c}=22\mu eV$ and where $R=V/I$ is the resistance of the contact. At higher temperature, the general behavior is rounded by thermal fluctuations.
Our noise results show two unexpected features: first the crossover to $S=2eI$ is predicted for voltage near the gap in the zero temperature models. As said before, in real superconducting-semiconducting junctions exhibiting the reflectionless tunneling a large quasiparticles contribution exists much below the gap. This component explains the observed crossover in accordance with our  estimations from conductance measurements, which indicate an increase of the quasiparticles contribution with voltage due to the increase of $T_{eff}$ in the silicon layer which is a bad thermal reservoir \cite{Quirion00}. The increase of $T_{eff}$  could also explain the absence of thermal crossover near  $V \simeq  5 \mu V$ at $T=50mK$, our second unexpected result. However the voltage dependent Johnson noise is rapidly overpassed by the shot noise which grows linearly with the current. The comparison with the noise characteristics at $T=100mK$ shows actually that  $T_{eff}$, which depends in a complicated way on electrical power dissipation via electrons, phonons and superconducting contacts, increases moderately. 
\par In one dimension, Beenakker et al. \cite{Beenakker97} calculated the zero energy conductance and shot noise in S-I-N diffusive junctions for various values of $\Gamma L/l$ which corresponds to the ratio of the resistance of the normal metal to the barrier resistance ($\Gamma$ is the barrier transparency, $L$ the length of the normal metal and $l$ its elastic mean free path). It is shown that, when $\Gamma \ll 1$, the shot noise reaches twice the full shot noise for $\Gamma L/l \lesssim 0.1$. At finite energy, Hekking et al. \cite{Hekking93} have shown that the length $L$ should be replaced by the phase coherence length $L_{\Phi}$.  From ref \cite{Quirion00} we get $0.25$ for the ratio $\Gamma L_{\Phi}/l$ with $\Gamma=3.5\, 10^{-2}$, $L_{\Phi}= 50 nm$ and $l=7 nm$. This estimate is in good enough agreement with Beenakker's predictions since we do not know precisely the properties of the silicon underneath the contact\cite{Quirion00}.   
\par In conclusion, we measured the shot noise in a S-I-N junction where a strongly disordered metal (highly degenerate silicon) is in contact with a superconductor (TiN). We found that the shot noise is equal to twice the full shot noise ($S=4eI$) at low energy in a regime where electron-hole coherence enhances the conductance. This result is expected from the theory \cite{Beenakker97} and corresponds to the Walter Schottky experiment \cite{Schottky} with field emission of Cooper pairs through a dielectrics. Electron-hole coherence is required to restore  a large enough $2e$-(Andreev) component of the subgap current allowing the measurement of the doubled charge. Above $20 \mu V$, the shot noise follows the Poisson noise $S_{Poisson}=2eI$ due to a dominant quasiparticle contribution. 
\par We thank P. Samuelsson,  M. B\"uttiker and X. Jehl for stimulating discussions and J.L. Thomassin for technical assistance.


\begin{thebibliography}{99}

\bibitem{Schottky} Schottky W., Ann. Phys. {\bf 57}, 541 (1918)
\bibitem{Blanter00} For a review on ``shot noise in mesoscopic conductors'' see Y. Blanter and M. B\"uttiker, Physics Reports {\bf 336}, (2000).
\bibitem{Quirion01} D. Quirion, C. Hoffmann, F. Lefloch and M. Sanquer, Phys. Rev. B {\bf 65}, 100508 (2002).
\bibitem{Kastalsky91} A. Kastalsky, A.W. Kleinsasser, L.H. Greene, R. Bhat, F.P. Milliken, J.P. Harbison, Phys. Rev. Lett. {\bf 67}, 3026 (1991).
\bibitem{VanWees92} B.J. Van Wees, P. de Vries, P. Magn\'ee and T.M. Klapwijk, Phys. Rev. Lett. {\bf 69}, 510 (1992).
\bibitem{Quirion00} D. Quirion, F. Lefloch, M. Sanquer, J. Low. Temp. Phys. {\bf 120}, 361 (2000).
\bibitem{Beenakker97} C. W. J. Beenakker, Review of Modern Phys. {\bf 69}, 731 (1997) and references therein.
\bibitem{Volkov94} A.F. Volkov, Physica B {\bf 203}, 267 (1994).
\bibitem{Bakker}S. J. M. Bakker et al., Phys. Rev. B {\bf 49}, 13275  (1994).
\bibitem{Magnee}P.H.C. Magn\'ee, N. van der Post, P. H. M. Kooistra, B. J. van Wees, and T. M. Klapwijk, Phys. Rev. B {\bf 50}, 4594  (1994).
\bibitem{Beltram}F. Giazotto et al., Appl. Phys. Lett. {\bf 78}, 1772 (2001).
\bibitem{Kozhevnikov00} A.A. Kozhevnikov, R.J. Schoelkopf, L.E. Calvet, D.E. Prober and M.J. Rooks, J. Low Temp. Phys. {\bf 118}, 671 (2000),A.A. Kozhevnikov, R.J. Schoelkopf, D.E. Prober, Phys. Rev. Lett. {\bf 84}, 3398 (2000). 
\bibitem{Jehl00} X. Jehl, M. Sanquer, R. Calemczuk and D. Mailly, Nature {\bf 405}, 50 (2000), X. Jehl and M. Sanquer, Phys. Rev. B {\bf 63}, 052511 (2001).
\bibitem{Dynes78} R. C. Dynes, V. Narayanamurti, and J. P. Garno, Phys. Rev. Lett. {\bf 41}, 1509 (1978).
\bibitem{Nagaev01} K.E. Nagaev and M. B\"uttiker, Phys. Rev. B {\bf 63}, 081301 (2001).
\bibitem{BTK82} G.E. Blonder, M. Tinkham and T.M. Klapwijk, Phys. Rev. B {\bf 25}, 4515 (1982).
\bibitem{Jehl99} X. Jehl, P. Payet-Burin, C. Baraduc, R. Calemczuk and M. Sanquer, Rev. Sci. Instrum. {\bf 70}, 2711 (1999).
\bibitem{Hekking93} F.W. Hekking and Y. Nazarov, Phys. Rev. Lett. {\bf 71}, 1625 (1993).


\end{thebibliography}
\end{document}